\documentclass[aps,prd,twocolumn,epsf]{revtex4}

\usepackage{graphicx}
\usepackage{amsmath,amssymb,latexsym}
\usepackage{bm}

\def\lbldef#1#2{\expandafter\gdef\csname #1\endcsname {#2}}
\def\eqn#1#2{\lbldef{#1}{(\ref{#1})}%
\begin{equation} #2 \label{#1} \end{equation}}

\def\href#1#2{#2}

\newcommand{\ber}{\begin{eqnarray}}
\newcommand{\eer}{\end{eqnarray}}

\newcommand{\beqar}{\begin{eqnarray}}

\newcommand{\eeqar}{\end{eqnarray}}

\newcommand{\dsl}
  {\kern.06em\hbox{\raise.15ex\hbox{$/$}\kern-.56em\hbox{$\partial$}}}

\newcommand{\eeqarr}{\end{eqnarray}}
\newcommand{\ZZ}{{\rm \kern 0.275em Z \kern -0.92em Z}\;}

\begin{document}

\preprint{
\hfil
\begin{minipage}[t]{3in}
\begin{flushright}
\vspace*{.4in}
NUB--xxx--Th--04\\
hep-ph/0504260
\end{flushright}
\end{minipage}
}

\title{Gluino Air Showers as a Signal of Split Supersymmetry}

\author{Javier G. Gonzalez, Stephen Reucroft, and John Swain}
\affiliation{Department of Physics,
Northeastern University, Boston, MA 02115
}

\begin{abstract}

It has been proposed recently that, within the framework of split 
Supersymmetry, long lived gluinos generated in astrophysical sources 
could be detected  using the signatures of the air showers they produce, 
thus providing a lower bound for their lifetime and for the scale of 
SUSY breaking. We present the longitudinal profile and lateral spread of 
$G$-hadron induced extensive air showers and consider the possibility of 
measuring them with a detector with the characteristics of the Pierre Auger Observatory.

\end{abstract}


\maketitle

\section{Introduction}

In the almost structureless fast falling with energy inclusive 
cosmic ray spectrum, three kinematic features have drawn considerable 
attention for a long time. These features, known as 
the knee, the ankle, and the ultraviolet cutoff, are the only ones in which 
the spectral index shows a sharper variation as a function of energy, 
probably signaling some ``new physics''. In particular, if cosmic ray sources are at 
cosmological distances, the cutoff is expected at about $10^{10.9}$~GeV, 
due to the GZK~\cite{Greisen:1966jv} interactions of the primaries with the microwave background radiation. 
The existence of data beyond the GZK cutoff~\cite{Takeda:1998ps} has been puzzling theorists and 
experimentalists~\cite{Abbasi:2002ta}, but a clear and widely accepted explanation is 
yet to see the light of day.
The proposed resolution for this puzzle generally invokes physics from the most favored theories 
beyond the standard model (SM) like string/M theory, supersymmetry (SUSY), 
grand unified theories (GUTs), and TeV-scale gravity~\cite{Anchordoqui:2002hs}.

A novel beyond--SM--model proposal to break the GZK barrier is to assume that 
ultrahigh energy cosmic rays are not known particles but a new species of particle, 
generally referred to as the uhecron, $U$~\cite{Farrar:1996rg}. The meager information we have 
about super-GZK particles allows a na\"{\i}ve description of the properties 
of the $U$. The muonic content in the atmospheric cascades suggests $U$'s should interact strongly. 
At the same time, if $U$'s are produced at cosmological distances, they 
must be stable, or at least remarkably long lived, with mean-lifetime 
$\tau \gtrsim 10^6 \, (m_U/3~{\rm GeV})\, (d/ {\rm Gpc})\,{\rm s},$ 
where $d$ is the distance to the source and $m_U$, the uhecron's mass.
Additionally, since the threshold energy increases linearly with $m_U$,
to avoid photo-pion production on the CMB $m_U \gtrsim 1.5$~GeV. Within the  
Minimal Supersymmetric (MS) extension of the SM, the allowed range for gluino masses is 
$m_{\tilde{g}} \leq 3~{\rm GeV}$ and $25~{\rm GeV} \leq m_{\tilde{g}} \leq 35~{\rm GeV}$. 
In this direction, it was noted in~\cite{Berezinsky:2001fy} that light  Supersymmetric 
baryons (made from a light gluino + the 
usual quarks and gluons, $m_U \lesssim 3$~GeV) would produce atmospheric cascades very similar to 
those initiated by protons.

Recently, Arkani-Hamed and Dimopoulos~\cite{Arkani-Hamed:2004fb} proposed an alternative 
to the MSSM in 
which the mass spectrum of the super-partners is split in two. In this theory, all the scalars, 
except for a 
fine tuned Higgs, get a mass at a high scale of supersymmetry breaking while the fermion's masses 
remain near the electroweak scale protected by chiral symmetry. Additionally, 
all corrections that involve loops of supersymmetric bosons are suppressed, thus removing most of the 
tunings required to reproduce $(g-2)_\mu$, $B-\bar{B}$ mixing and 
$b \rightarrow s\gamma$~\cite{Giudice:2004ss}. At the same 
time it allows for radiative corrections to the Higgs mass. Moreover, very recently it was 
shown that there exists a realization of a ``split SUSY'' in String 
Theory~\cite{Antoniadis:2004dt,Arkani-Hamed:2004ss}.

An important feature of split SUSY 
is the long life of the gluino due to the high masses of the virtual scalars ($m_s$) that mediate 
the decay. Indeed, very strong limits on heavy isotope abundance require the gluino to decay on Gyr 
time scales, leading to an upper bound for the scale of SUSY breaking ${\cal O} (10^{13})$~GeV.
Additionally, it has been pointed out that the detection of gluinos coming from 
astrophysical sources (with $m_{\tilde g} \sim 500$~GeV) leads to a lower bound on their proper 
lifetime of the order of 100 yr, which translates into a lower  bound on the scale 
of SUSY breaking, ${\cal O} (10^{11})$~GeV~\cite{Anchordoqui:2004bd}. 

In light of this, it is of interest to explore the potential of forthcoming cosmic ray 
observatories to observe gluino-induced events. Some signatures of the air showers initiated by 
these long lived gluinos have been 
presented in~\cite{Anchordoqui:2004bd,Hewett:2004nw}. In this Brief Report, we carry out a more 
detailed analysis by generating gluino air showers through Monte Carlo simulations and pave the ground 
for a future study on the actual feasibility of measuring them at the Pierre Auger 
Observatory~\cite{Abraham:2004dt}. The outline 
is as follows. In Sec. II we review the relevant aspects of cosmic ray air showers. 
In Sec. III we first carry out Monte Carlo simulations of gluino induced showers 
and then show their distinct signatures in the air shower profile and lateral spread at ground level.
Section IV contains a summary of our results.

\section{Cosmic Ray Air Showers}

When a high energy particle hits the atmosphere it generates a roughly conical cascade of secondary 
particles, an air shower. At any given time, the shower can be pictured as a bunch of particles, 
the shower front, traveling toward the ground at nearly the speed of light. The number of particles 
in the shower multiplies as the front traverses the atmosphere, until the particles' energy fall below 
a threshold at which ionization losses dominate over particle creation, after this point the number 
of particles decreases. By the time the front hits the ground, its shape is similar to that of 
a ``saucer'' with a radius that can range from a few meters to a few kilometers. The shape of the 
shower front is actually closer to that of a spherical shell, with a curvature of a few kilometers 
for almost vertical showers to more than a hundred kilometers for inclined ones. The number of particles 
as a function of amount of atmosphere traversed is the ``longitudinal profile'' of the shower.

The general properties of the longitudinal profile can be understood with a simple model and it usually 
can be parameterized by a Gaisser-Hillas function~\cite{Anchordoqui:2004xb},
\eqn{gaiserhillas}{N_e\left(X\right) = N_{e,{\rm max}}\left(\frac{X-X_0}{X_{{\rm max}}-X_0} 
\right)^{\frac{X-X_{\rm max}}{\lambda}}e^{\frac{X-X_{\rm max}}{\lambda}}}
where $N_{e, {\rm max}}$ is the number of particles at shower maximum, $X_0$ is the depth of the first 
observed interaction, $X_{\rm max}$ is the depth at the maximum, and $\lambda = 70$~g/cm$^2$. The position 
of the maximum, $X_{\rm max}$, depends on the energy as well as on the nature of the primary particle. 
With cosmic ray showers however, there are fluctuations, associated mainly with the point where the 
primary first interacts, as well as statistical fluctuations in the development of the shower.

When the shower front reaches the ground it is spread over an area of up to a few kilometers. It is then 
possible to study the density of energy deposited (or particle densities), on the ground as a function 
of time and position. There are a number of parameterizations for such distributions but they are mostly 
modified power laws like the following~\cite{Anchordoqui:2004xb}
\eqn{NKG}{S\left(r\right) = C \left(\frac{r}{r_M}\right)^{-\alpha} 
\left(1+\frac{r}{r_M}\right)^{-\eta+\alpha}}
where $r$ is the distance to the point where the core hits the ground, $r_M$ is the Moliere radius at 
two radiation lengths above the observation level, $\alpha$ is another empirical parameter and $\eta$ 
is the free parameter that depends on the angle. The normalization constant $C$ will depend on the energy.

If the primary particle is a hadron, the first interaction will be a hadronic interaction and the number 
of hadrons, mostly pions, will increase with each interaction. Statistically, in each interaction, about 
30\% of the energy goes into neutral pions that decay, generating an electromagnetic cascade. 
In this way, the energy of the primary is split into an electromagnetic part and a ``muonic'' part, 
that comes from the $\pi^{\pm}$ that managed to decay. If the incoming primary has a higher energy, 
the number of interactions required to lower the energy per particle under the threshold at which the 
pions will most likely decay increases, increasing the fraction of the energy that goes into the 
electromagnetic part. As a result, the number of muons in a shower scales as $E^{0.94}$. This in turn 
implies that the number of muons for a nucleus of mass A relates to the number of muons on a proton 
shower: $N_{\mu}^A = A^{0.06}N_{\mu}^{p}$ \cite{Anchordoqui:2004xb}. At this point it is worth 
noting that these numbers are strongly dependent on the particular hadronic interaction model used. 
There are three hadronic interaction models commonly used in air shower 
simulations, {\sc sibyll}~\cite{Fletcher:1994bd}, {\sc qgsjet}~\cite{Kalmykov:1997te} and 
{\sc dpmjet}~\cite{Ranft:1994fd}. 
All of them are extrapolations of 
models that agree with the experimental data but show a different behaviour at energies 
beyond 10 TeV center of mass energy~\cite{Anchordoqui:1998nq}.

If the primary particle is a $\gamma$-ray the interactions that occur are mostly pair production, 
Bremsstrahlung, ionization losses and Compton scattering. Also, at energies higher than $10^{10}$~GeV, 
the LPM effect suppresses the cross-section for pair production and Bremsstrahlung. This results in the 
shower being more elongated.

\section{Gluino air showers}

In this section we will study gluino induced showers.
To carry out this study we use {\sc aires}, a set of programs specifically 
designed to simulate the extensive air showers
generated by ultra high energy cosmic rays interacting with the atmosphere.
The {\sc Aires} system is described elsewhere ~\cite{Sciutto:1999,Sciutto:2001dn} and takes into account the 
relevant interactions, including electromagnetic and hadronic interactions and transport processes. The 
hadronic model used is {\sc Sibyll}.

We use a feature of {\sc aires} that allows for the definition of special primaries by providing 
a program that handles the first interactions of each  primary until the main 
program ({\sc aires}) can take over and simulate the rest of the shower until it strikes ground. 
To model the gluino induced showers we first determine where there will be a major 
interaction and then inject a 
proton with energy equal to the energy deposited by the gluon. We then force each proton to 
have it's first interaction
at its corresponding point, giving rise to a hadronic shower that is then simulated by the 
standard \cite{Sciutto:1999,Sciutto:2001dn} program.

The gluino containing hadron (hereafter $G$) cross section is about half the pion-air cross 
section and the inelasticity is $K \propto 1~{\rm GeV}/M_G$~\cite{Anchordoqui:2004bd,Berezinsky:2001fy}. 
The gluino mass range is not constrained so we could, in principle, study it at different scales. 
According to~\cite{Hewett:2004nw}, the masses accesible to neutrino detectors are $\leq 170$~GeV. 
In our case we try to probe for higher masses, while keeping the fluxes within reach. In our simulations 
we adopt a gluino mass of $500$~GeV, which yields an inelasticity of $0.002$~\cite{Anchordoqui:2004bd}.
This small inelasticity is precisely what allows one to model a $G$ shower as a series of proton 
sub-showers separated according to the $G$ mean
free path, with each proton having about $0.002$ of the original $G$ energy. 

In our case, this program takes the $G$-hadron of a given energy, mean free path and inelasticity.  
The nature of the particle chosen to be injected at each vertex will determine the amount of energy 
channeled into the hadronic shower and the 
amount of energy going into the electromagnetic shower. For a detector like the surface array of 
the Pierre Auger Observatory, 
the hadronic part (the muons) are enhanced over the electromagnetic part 
(also considering a great part of the electromagnetic part has died away in flight). 
This means that, by injecting a proton in each vertex we are underestimating the number of muons 
that can be sampled in the ground. 

The proton sub-showers are generated following these simple steps, and are repeated until the $G$-hadron 
reaches ground level:
\begin{itemize}
\item Calculate the point of the next interaction.
\item Decrease the energy of the G-hadron by a factor given by the inelasticity.
\item Inject a proton with energy equal to the decrease in the G-hadron's energy traveling in 
the same direction.
\end{itemize}

Once the initial string of protons is generated, the {\sc aires} package
takes over simulation of the standard physics interactions and transport through
the atmosphere to produce the set of showers.

\begin{figure}
\includegraphics[width=0.37\textwidth]{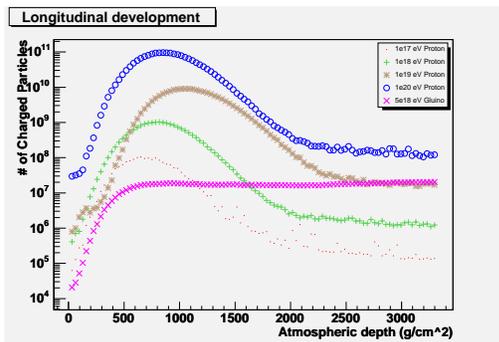}
\caption{\label{fig:comparison} Longitudinal profile for protons at  $10^{17}$, $10^{18}$, $10^{19}$, $10^{20}$ eV and a gluino at $5\times10^{18}$ eV.}
\end{figure}

In Fig.~\ref{fig:comparison} we present the average longitudinal profile for
100 $G$-hadron induced air showers, along with the longitudinal development of a
proton with different energies. Our shower simulations have been carried out for incident 
zenith angle $75^\circ$. It should be clear that the development of a
$G$-hadron shower can not be fitted by the Gaisser-Hillas function given in Eq.~(\ref{gaiserhillas}).
One of the biggest sources of fluctuations in air showers is the first interaction point. 
This means that, in our case, the longitudinal development of any particular shower will show 
small fluctuations in the shape, since one of these is composed of about ten proton showers.  
It should be noted that it might be possible 
to isolate these events from their background even for zenith angles as low as $60^\circ$, 
where the atmospheric depth is around 2000 g/cm, still more than three times the point where 
the shower reaches its maximum.

\begin{figure}
\includegraphics[width=0.37\textwidth]{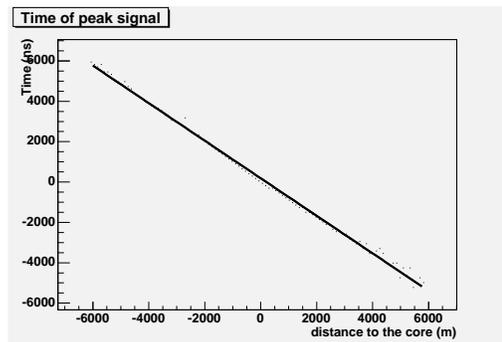}
\caption{\label{fig:timeProfile} Time of arrival versus distance to the core, 
along the symmetry axis of the LDF}
\end{figure}

In Fig.~\ref{fig:timeProfile} we show the arrival time as a function of distance to the core 
along the major symmetry axis of the lateral distribution function, 
again for an average over 100 showers. A fit to this plot, taking the spherical front approximation 
to determine the radius of curvature of the shower front, leads to a value around 78~km. In the 
general case of cosmic ray air showers the front is not parameterized as a sphere since the curvature 
is more pronounced near the core and flattens out for big radius.

In the spherical front approximation, the arrival time as a function 
of the distance to the core ($r$) is, up to terms of order greater than $r/R$
\begin{equation}
t(r)= - \frac{r}{c} \hat{u}_r \cdot \hat{u}_R + \frac{r^2}{cR}\left(1 - \left(\hat{u}_r \cdot 
\hat{u}_R\right)^2\right) \,\,,
\end{equation}
where $R$ is the radius of the front and $\hat{u}_R$ is the unit vector
pointing in the direction of arrival of the shower and $\hat{u}_r$ is the
radial unitary vector on the detector plane.

\begin{figure}
\includegraphics[width=0.37\textwidth]{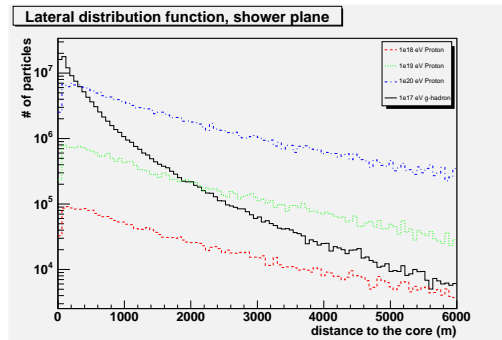}
\caption{\label{fig:ldf} Average Lateral spread for  $10^{18}$, $10^{19}$, $10^{20}$ eV  proton showers 
and $5\times 10^{18}$ eV gluino showers.}
\end{figure}

The lateral distribution function (LDF) for all particles also shows a distinct behavior. 
In Fig.~\ref{fig:ldf} the total LDF on the shower plane is plotted, along with the corresponding 
ones for protons of different energies. A distinct feature of the LDF from $G$-hadrons is that the 
slope depends on the distance to the core, as opposed to proton generated ones. It is due to the 
fact that $G$-hadron induced showers are just a superposition of lower energy showers with different 
ages (hence different slopes). The younger showers, that are spread over a smaller area, show a steeper 
LDF. The particle densities are on the order of the densities from proton showers. These densities need to 
be scaled according to the detector response. Also, for very inclined showers, the steepness of the LDF 
depends strongly on the zenith angle so a search for these has to take into account the angular resolution 
of the experiment.

\section{Summary}

Using a simple model to simulate the interaction of $G$-hadrons hitting the top of the atmosphere, 
we have shown that the longitudinal development of a $G$-hadron induced air shower is very distinct 
from that of protons and gamma rays that compose the main background. Thanks to the low inelasticity 
of $G$-air interactions, the $G$-hadron will produce a sequence of smaller showers of almost the 
same energy, and the value for the cross-section, in agreement with previous works, gives 
a separation between them that makes it impossible to resolve them.

It should then be possible to differentiate between $G$-hadron induced showers and those of the 
background, since they display such a unique profile. In order to assess how difficult it will be to 
actually measure these showers we need to consider the characteristics of the detector. In the case of 
fluorescence techniques, a careful study of the signal to noise ratio for the intensities associated 
with these showers will be needed in order to correctly estimate the correct aperture.

It was also shown that the lateral distribution of particles at ground level is such that it should be 
measurable with a ground array. The expected signature, a small front radius given by the arrival times, 
seems not to be realized but the LDF does show a varying slope, due to the superposition of different 
age showers. A more detailed study also needs to be done in order to study the aperture and the 
discrimination power, since these will depend strongly on the characteristics of the detector.\\

\underline{Note added:} After this paper was finished new bounds on the $M_G$-SUSY 
breaking scale plane were reported~\cite{Arvanitaki:2005fa}. Interestingly, a 
small window for high scale SUSY breaking and $M_G = 500$~GeV still remains open.

\section*{Acknowledgements}

We would like to thank Luis Anchordoqui and Carlos Nu\~{n}ez for usefull discussions.
This work has been partially supported by the NSF grant No. PHY-0140407.


\begin{thebibliography}{99}


\bibitem{Greisen:1966jv}
K.~Greisen,
Phys.\ Rev.\ Lett.\  {\bf 16}, 748 (1966);
G.~T.~Zatsepin and V.~A.~Kuzmin,
JETP Lett.\  {\bf 4}, 78 (1966)
[Pisma Zh.\ Eksp.\ Teor.\ Fiz.\  {\bf 4}, 114 (1966)].

\bibitem{Takeda:1998ps}
M.~Takeda {\it et al.},
Phys.\ Rev.\ Lett.\  {\bf 81}, 1163 (1998)
[arXiv:astro-ph/9807193].


\bibitem{Abbasi:2002ta}
It should be stressed that the most recent results reported 
by the HiRes Collaboration describe a 
spectrum which is consistent with the expected GZK feature. 
R.~U.~Abbasi {\it et al.}  [HiRes Collaboration],
Phys.\ Rev.\ Lett.\  {\bf 92}, 151101 (2004)
[arXiv:astro-ph/0208243].


\bibitem{Anchordoqui:2002hs} For a review, see {\it e.g},
L.~Anchordoqui, T.~Paul, S.~Reucroft and J.~Swain,
Int.\ J.\ Mod.\ Phys.\ A {\bf 18}, 2229 (2003)
[arXiv:hep-ph/0206072].

\bibitem{Farrar:1996rg}
G.~R.~Farrar,
Phys.\ Rev.\ Lett.\  {\bf 76}, 4111 (1996)
[arXiv:hep-ph/9603271].

\bibitem{Berezinsky:2001fy}
V.~Berezinsky, M.~Kachelriess and S.~Ostapchenko,
Phys.\ Rev.\ D {\bf 65}, 083004 (2002)
[arXiv:astro-ph/0109026].



\bibitem{Arkani-Hamed:2004fb}
N.~Arkani-Hamed and S.~Dimopoulos,
arXiv:hep-th/0405159.
%

\bibitem{Giudice:2004ss}
G.F. ~Giudice and ~A. Romanino
arXiv:hep-ph/0406088.


\bibitem{Antoniadis:2004dt}
I.~Antoniadis and S.~Dimopoulos,
arXiv:hep-th/0411032;
B.~Kors and P.~Nath,
arXiv:hep-th/0411201.


\bibitem{Arkani-Hamed:2004ss} For general aspects of split SUSY, see {\it e.g.,}
N.~Arkani-Hamed, S.~Dimopoulos, G.F. Giudice and A. Romanino
arXiv:hep-ph/0409232.



\bibitem{Anchordoqui:2004bd}
L.~Anchordoqui, H.~Goldberg and C.~Nunez,
arXiv:hep-ph/0408284.



\bibitem{Hewett:2004nw}
J.~L.~Hewett, B.~Lillie, M.~Masip and T.~G.~Rizzo,
JHEP {\bf 0409}, 070 (2004)
[arXiv:hep-ph/0408248].

\bibitem{Abraham:2004dt}
J.~Abraham {\it et al.}  [Pierre Auger Collaboration],
Nucl.\ Instrum.\ Meth.\ A {\bf 523} (2004) 50.



\bibitem{Anchordoqui:2004xb}
L.~Anchordoqui, M.~T.~Dova, A.~Mariazzi, T.~McCauley, T.~Paul, S.~Reucroft and J.~Swain,
Annals Phys.\  {\bf 314}, 145 (2004)
[arXiv:hep-ph/0407020].

\bibitem{Fletcher:1994bd}
  R.~S.~Fletcher, T.~K.~Gaisser, P.~Lipari and T.~Stanev,
  Phys.\ Rev.\ D {\bf 50}, 5710 (1994).
R. ~Engel, T.K ~Gaisser and T. ~Stanev,
Proc. 26th ICRC (Utah), {\bf 1}, 415 (1999).

\bibitem{Kalmykov:1997te}
  N.~N.~Kalmykov, S.~S.~Ostapchenko and A.~I.~Pavlov,
  Nucl.\ Phys.\ Proc.\ Suppl.\  {\bf 52B}, 17 (1997).

\bibitem{Ranft:1994fd}
  J.~Ranft,
  Phys.\ Rev.\ D {\bf 51}, 64 (1995).


\bibitem{Anchordoqui:1998nq}
  L.~A.~Anchordoqui, M.~T.~Dova, L.~N.~Epele and S.~J.~Sciutto,
  Phys.\ Rev.\ D {\bf 59}, 094003 (1999)
  [arXiv:hep-ph/9810384].





\bibitem{Sciutto:2001dn}
S.~J.~Sciutto,
arXiv:astro-ph/0106044.

\bibitem{Sciutto:1999}
S.~J.~Sciutto,
arXiv:astro-ph/9911331.



\bibitem{Arvanitaki:2005fa}
  A.~Arvanitaki, C.~Davis, P.~W.~Graham, A.~Pierce and J.~G.~Wacker,
  arXiv:hep-ph/0504210.



\end{thebibliography}
\end{document}